\documentclass[journal=jcplcd,manuscript=letter]{achemso}
\usepackage[version=3]{mhchem} % Formula subscripts using \ce{}
\usepackage{siunitx}

\usepackage{amssymb}
\usepackage{graphicx}
\graphicspath{{./images/}}
\usepackage{bm}
\usepackage[colorlinks=true,linkcolor=black]{hyperref}
\usepackage{color}
\usepackage{ulem}

\author{Sven Bodenstedt}
\affiliation{ICFO-Institut de Ci\`encies Fot\`oniques, The Barcelona Institute of Science and Technology, 08860 Castelldefels (Barcelona), Spain}
\author{Denis Moll}
\affiliation{NMR Signal Enhancement Group, Max Planck Institute for Biophysical Chemistry, 37077 G{\"o}ttingen, Germany}
\altaffiliation{Center for Biostructural Imaging of Neurodegeneration, UMG, 37075 G{\"o}ttingen, Germany}
\author{Stefan Gl{\"o}ggler}
\affiliation{NMR Signal Enhancement Group, Max Planck Institute for Biophysical Chemistry, 37077 G{\"o}ttingen, Germany}
\altaffiliation{Center for Biostructural Imaging of Neurodegeneration, UMG, 37075 G{\"o}ttingen, Germany}
\author{Morgan W. Mitchell}
\affiliation{ICFO-Institut de Ci\`encies Fot\`oniques, The Barcelona Institute of Science and Technology, 08860 Castelldefels (Barcelona), Spain}
\altaffiliation{ICREA -- Instituci\'{o} Catalana de Recerca i Estudis Avan\c{c}ats, 08010 Barcelona, Spain}
\author{Michael C. D. Tayler}
\email{michael.tayler@icfo.eu}
\affiliation{ICFO-Institut de Ci\`encies Fot\`oniques, The Barcelona Institute of Science and Technology, 08860 Castelldefels (Barcelona), Spain}

\title{Decoupling of Spin Decoherence Paths near Zero Magnetic Field}
\abbreviations{NMR}
\keywords{Nuclear magnetism, Spin relaxation, Zero-field NMR, Magnetometry}
\begin{document}

%%%%%%%%%%%%%%%%%%%%%%%%%%%%%%%%%%%%%%%%%%%%%%%%%%%%%%%%%%%%%%%%%%%%%
%% The "tocentry" environment can be used to create an entry for the
%% graphical table of contents. It is given here as some journals
%% require that it is printed as part of the abstract page. It will
%% be automatically moved as appropriate.
%%%%%%%%%%%%%%%%%%%%%%%%%%%%%%%%%%%%%%%%%%%%%%%%%%%%%%%%%%%%%%%%%%%%%
\begin{tocentry}
\centering
\includegraphics[width = 4.5cm]{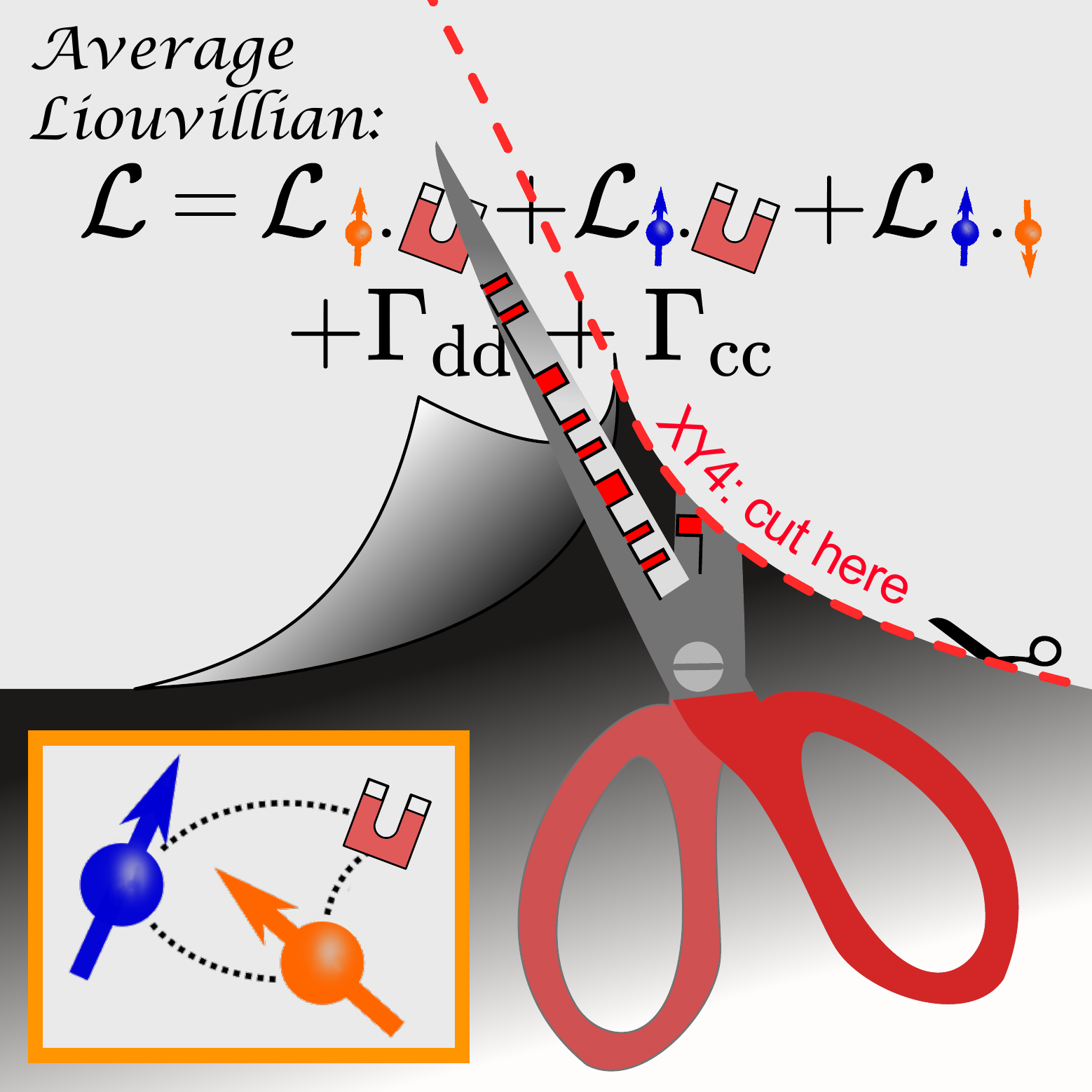}
\end{tocentry}

%%%%%%%%%%%%%%%%%%%%%%%%%%%%%%%%%%%%%%%%%%%%%%%%%%%%%%%%%%%%%%%%%%%%%
%% The abstract environment will automatically gobble the contents
%% if an abstract is not used by the target journal.
%%%%%%%%%%%%%%%%%%%%%%%%%%%%%%%%%%%%%%%%%%%%%%%%%%%%%%%%%%%%%%%%%%%%%
\begin{abstract}
We demonstrate a method to quantify and manipulate nuclear spin decoherence mechanisms that are active in zero to ultralow magnetic fields.  These include: (i) non-adiabatic switching of spin quantization axis, due to residual background fields; (ii) scalar pathways due to through-bond couplings between \textsuperscript{1}H and heteronuclear spin species, such as \textsuperscript{2}H used partially as an isotopic substitute for \textsuperscript{1}H.  Under conditions of free evolution, scalar relaxation due to \textsuperscript{2}H can significantly limit nuclear spin polarization lifetimes and thus the scope of magnetic resonance procedures near zero field.  It is shown that robust trains of pulsed dc magnetic fields that apply $\pi$ flip angles to one or multiple spin species may switch effective symmetry of the nuclear spin Hamiltonian, imposing decoupled or coupled dynamic regimes on demand.  The method should broaden the spectrum of hyperpolarized biomedical contrast-agent compounds and hyperpolarization procedures that are used near zero field.

\vspace{1cm}

\centering TOC graphic:\\
	\includegraphics[width=5.08cm]{toc.pdf}

\end{abstract}

%%%%%%%%%%%%%%%%%%%%%%%%%%%%%%%%%%%%%%%%%%%%%%%%%%%%%%%%%%%%%%%%%%%%%
%% Start the main part of the manuscript here.
%%%%%%%%%%%%%%%%%%%%%%%%%%%%%%%%%%%%%%%%%%%%%%%%%%%%%%%%%%%%%%%%%%%%%

\newpage
Thermodynamic behavior of a system is usually determined by internal degrees of freedom that depend on chemical and physical properties such as inter-particle couplings and symmetry.  While these may not be easily changed, it is well known (but sometimes overlooked) that the effect of internal properties on system dynamics can be controlled by manipulating the symmetry of the external environment.  For instance, dynamical decoupling\cite{Souza2012PhilTransRoyalSocA,Levitt1992PRL69,Levitt1994BMR16,Ghose1999MolPhys96}, spin locking\cite{Carravetta2004JACS126,Kimmichbook}, magnetic field cycling\cite{Kimmichbook,Kimmich1979BMR1,Kimmich2004PNMRS} and other symmetry switching techniques\cite{Levitt2020Pileiobook} can influence the rates at which a spin ensemble relaxes to thermodynamic equilibrium.  These techniques allow the internal variables (such as spin dipole-dipole or spin-rotation couplings) to be probed, even though relaxation is characterized by local, random interactions that fluctuate extremely rapidly relative to coherent time scales.

In this Letter, we examine the nature of nuclear spin relaxation near zero magnetic field and its manipulation by dynamical decoupling pulse sequences.  For a given spin system, a near-zero-field regime is characterized by inter-spin couplings that are large in magnitude compared to couplings with external magnetic fields.  One appeal of zero field is the existence of long-lived heteronuclear spin coherences\cite{Emondts2014PRL112,Tayler2018JMR297}, which can be detected using magnetometers with high sensitivity in the sub-kHz band\cite{BudkerOpticalMagnetometrybook,Tayler2017RSI} enabling an ultrahigh resolution form of nuclear magnetic resonance (NMR) spectroscopy\cite{Zhukov2018PCCP,Kiryutin2020Pileiobook,Blanchard2016emagres,Blanchard2013JACS}. Another appeal is the efficiency of nuclear spin hyperpolarization techniques close to zero field, such as \textsuperscript{13}C polarization via parahydrogen \cite{Goldman2006CRC9,Cavallari2015JPCB119,Stewart2018JMR296}.  However, interference of other nuclei and especially those that possess a quadrupolar moment has shown to tremendously reduce polarization lifetimes and polarization transfer efficiency.\cite{Hartwig2011JCP135,Barskiy2017CPC,Tayler2019JMR298}

A dual-species system, $IS$, is analyzed in which relaxation of the $I$ species is slow compared to the strength of couplings to $S$ and also the self-relaxation of $S$.  The polarization lifetime of $I$ is thus limited by $S$, however, the limit may be removed by dynamically decoupling either spin from the other.  The scenario is experimentally applied to $I$ nuclei of spin quantum number 1/2 (hydrogen-1) that couple to $S$ nuclei with nonzero electric quadrupole moment (spin $\geq$1, namely nitrogen-14 or deuterium).

\begin{figure}[t]%
	\includegraphics[width=8cm]{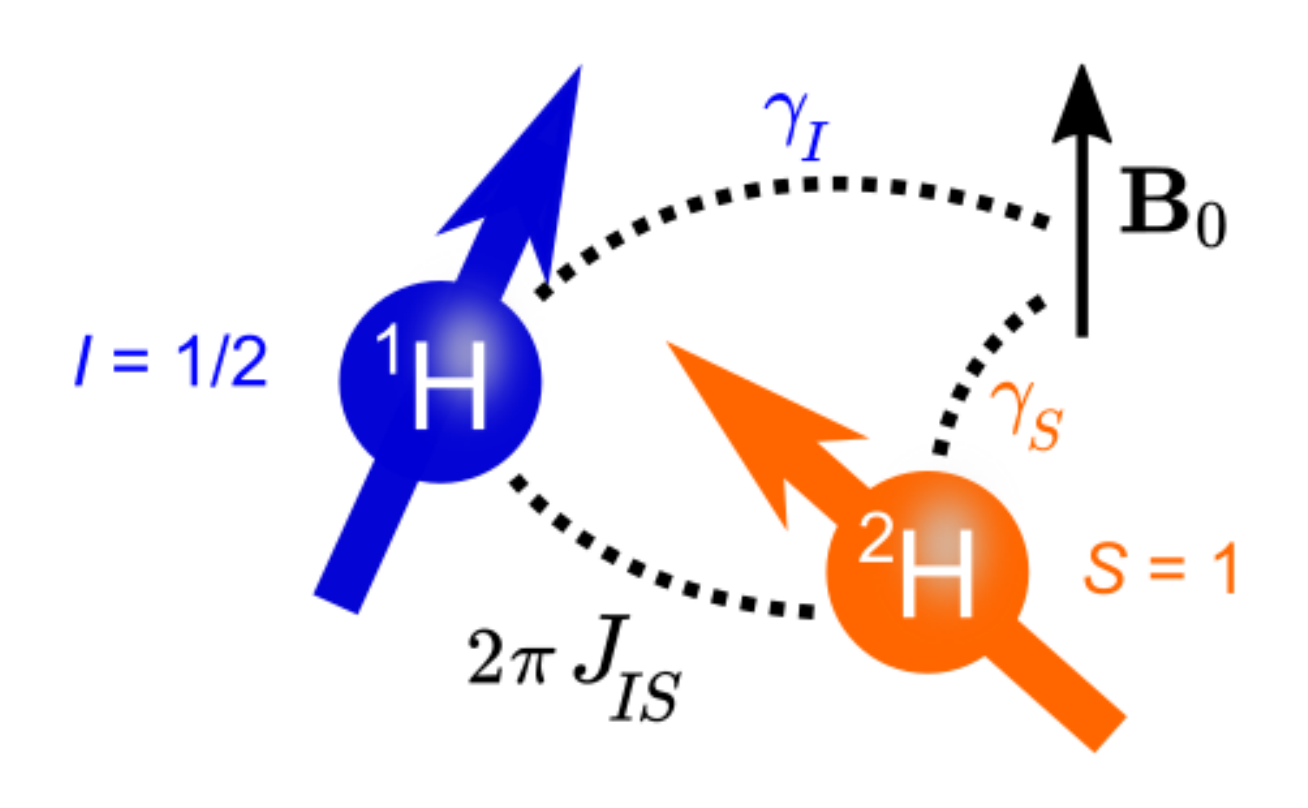}
	\caption{Model spin system comprising two coupled hydrogen nuclei $I$=\textsuperscript{1}H and $S$=\textsuperscript{2}H in a magnetic field $\mathbf{B}_0$.}% 
	\label{fig:IS}
\end{figure}

The following model is used to gain insight into spin symmetry switching near zero field.  An $IS$ pair (\autoref{fig:IS}) evolves freely via spin-spin and spin-field couplings given by a Hamiltonian
\begin{eqnarray}
\mathcal{H}(J_{IS},\mathbf{B}) &=& 2\pi J_{IS} (I_xS_x+I_yS_y+I_zS_z) \label{eq:HIS}  \\ 
&&- \gamma_I (I_xB_x+I_yB_y+I_zB_z) \nonumber \\
&&- \gamma_S (S_xB_x+S_yB_y+S_zB_z), \nonumber
\end{eqnarray}
in units of $\hbar$, where $J_{IS}$ is the strength of the coupling in Hz, $\mathbf{B}=\{B_x,B_y,B_z\}$ denotes the magnetic field vector, $\gamma_I$ and $\gamma_S$ are the gyromagnetic ratios in angular units.  The system is assumed to experience pulsed magnetic fields at regular time intervals $\tau$ to rotate the spin state instantaneously and periodically, such that  $\mathbf{B}(t) =\mathbf{B}_0 +  \sum_{j=1}^n \mathbf{B}_j \delta(t/\tau -j)$, where $\delta$ is the Dirac function, $\mathbf{B}_0$ is the constant bias field and $\{ \mathbf{B}_j \}$ are the field  pulse amplitudes.  The spin density operator $\rho$ after $n$ pulse-delay events is expressed mathematically as
\begin{equation}
    \rho(n\tau)=\prod_{j=1}^{n}R_j\exp{(\mathcal{L}_0?\tau)}\rho(0).
    \label{eq:rhot}
\end{equation}
Here superoperators $\{R_j\}$ denote each transformation under the pulsed fields, $\mathcal{L}_0$ is the Liouvillian superoperator defined by the master equation ${\rm d}\rho/{\rm d}t = \mathcal{L}_0\rho = \mathrm{i} [\rho\mathcal{H}(J_{IS},\mathbf{B}_0)-\mathcal{H}(J_{IS},\mathbf{B}_0)\rho]  - \Gamma \rho$ and $\Gamma$ is the relaxation superoperator.  For simplicity -- and because in the experiments later described the equilibrium polarization of the spin system is negligible compared to that of the initially prepared state -- the thermal equilibrium density operator is ignored.

Coherent averaging theory is used to transform the right-hand side of \autoref{eq:rhot} into a simple form $\rho(n\tau)=\exp{(\overline{\mathcal{L}}n\tau)}\rho(0)$, where $\overline{\mathcal{L}}$ is the \textit{average Liouvillian}.  The lowest-order terms of $\overline{\mathcal{L}}$
\begin{eqnarray}
\overline{\mathcal{L}} &=& \mathcal{L}^{(0)} + \mathcal{L}^{(1)} + \ldots\\
\mathcal{L}^{(0)} &=& (1/n) \sum_{j=1}\mathcal{L'}_j \nonumber \\
\mathcal{L}^{(1)} &=& (1/2n) \sum_{j=1,k<j}[\mathcal{L'}_j,\mathcal{L'}_k],\nonumber
\end{eqnarray}
with 
\begin{eqnarray}
    \mathcal{L}_j' &=& R_n R_{n-1}\ldots R_j \mathcal{L}_0 R_j^{-1}\ldots R_{n-1}^{-1} R_{n}^{-1},
\end{eqnarray}
are the most important to consider, but if $\tau^{-1}$ is fast compared to both $J_{IS}$ and $\gamma|\bm{B}|$ then $\overline{\mathcal{L}} = \mathcal{L}^{(0)}$ is a good approximation.  In this form the combined dynamic effects of $\mathcal{H}$ and $\Gamma$ may be analyzed.  We consider three distinct scenarios for $\{R_j\}$.  For sake of reference, the first is where all transformations $R_j$ equal an identity operation (no pulses, $\mathbf{B}_{j>0}=\mathbf{0}$).  The second is where pulses rotate the $I$ spin vector by $\pi$ radians alternately about $x$ and $y$ axes, leaving the $S$ spin vector unchanged: $\{R_j\} = \{ R_x^{(I)}(\pi)R_x^{(S)}(0),$ $ R_y^{(I)}(\pi)R_y^{(S)}(0),$ $ R_x^{(I)}(\pi)R_x^{(S)}(0),$ $ R_y^{(I)}(\pi)R_y^{(S)}(0), \ldots \}$.  This repeat unit of the pulse sequence is denoted as XY4($I$) and is the zero-field analog of the Carr-Purcell sequence for $I$\cite{Lee1987JMR75}.  Thirdly, both $I$ and $S$ are rotated by $\pi$ alternately about the $x$ and $y$ axes, which is denoted as XY4($I+S$): $\{R_j\} = \{ R_x^{(I)}(\pi)R_x^{(S)}(\pi),$ $ R_y^{(I)}(\pi)R_y^{(S)}(\pi),$ $ R_x^{(I)}(\pi)R_x^{(S)}(\pi),$ $ R_y^{(I)}(\pi)R_y^{(S)}(\pi), \ldots \}$.  It can be shown in the latter scenarios that $\mathcal{L}^{(0)}$ does not contain any spin operators that are linear in the rotation.  As an example, XY4($I+S$) eliminates all $\mathbf{B}_0$ terms from $\mathcal{L}^{(0)}$ meaning that the system behaves as if the field is zero.  In contrast, XY4($I$) averages $I_xS_x$, $I_yS_y$ and $I_zS_z$ to zero, so that the coupling between the spins appears to vanish.  

\begin{figure*}%
	\includegraphics[width=\columnwidth]{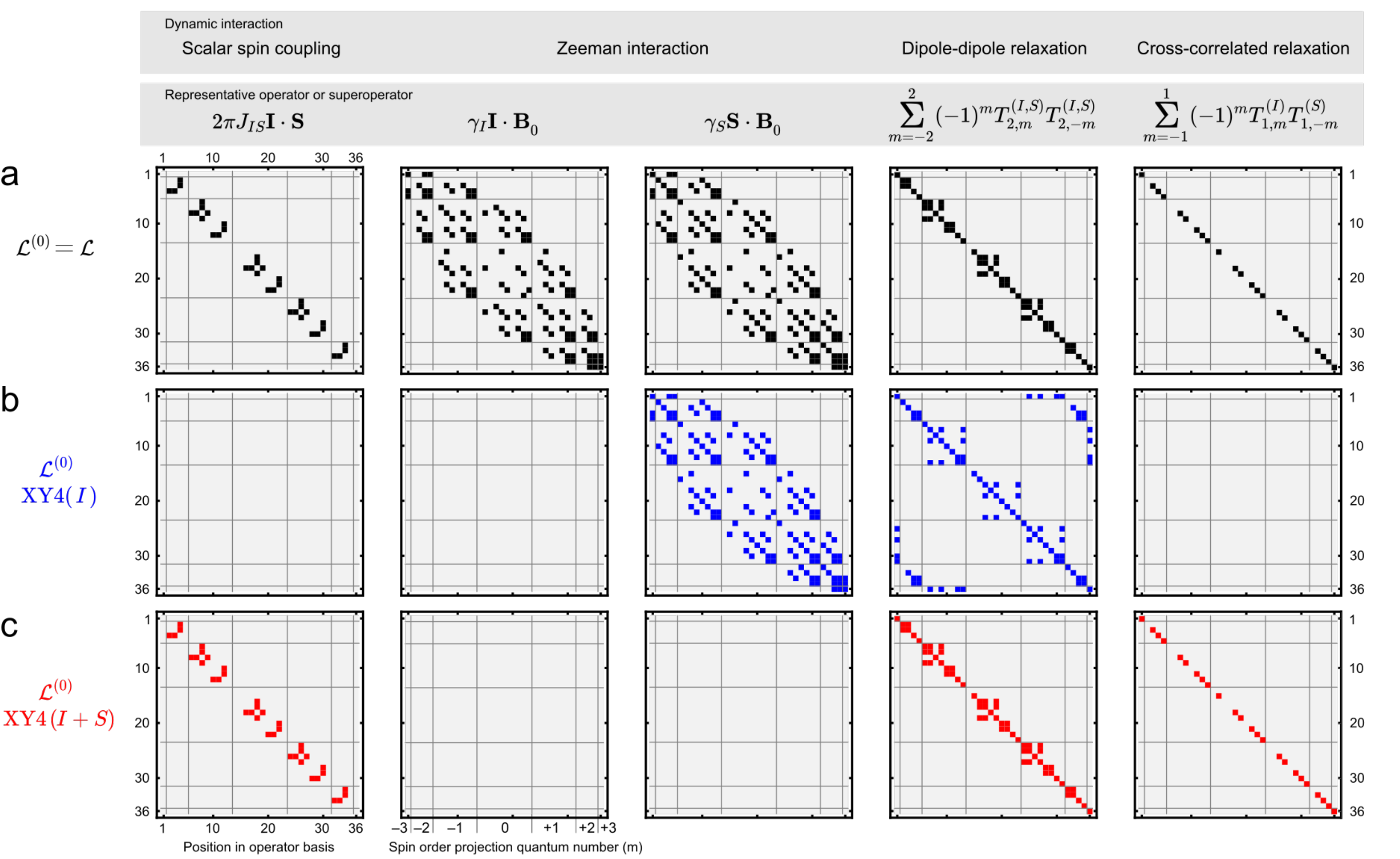}
	\caption{Symbolic matrix representations of $\mathcal{L}^{(0)}$ for interactions within a two-particle system comprising spin-1/2 (I) and spin-1 (S) species.  The representation is made in the spherical tensor operator basis sorted by projection index $m$ to illustrate block diagonal properties, where gray lines demarcate blocks of different $m$.  Shading corresponds to a nonzero matrix element of $\mathcal{L}^{(0)}$ between two basis operators, and the matrix dimension equals the number of basis operators $(2I+1)^2(2S+1)^2 = 36$.  Three scenarios are presented: (a) free evolution, i.e.\ $\mathcal{L}^{(0)}=\mathcal{L}_0$, (b) XY4 applied to the I spin, (c) XY4 applied to both I and S.  Analytical expressions for all matrix elements are provided in the Supporting Information.}% 
	\label{fig:Lmat}
\end{figure*}

\autoref{fig:Lmat} shows the effect of XY4 through matrix representations of contributions to $\mathcal{L}^{(0)}$.  Nonzero matrix elements are shaded corresponding to ${\rm tr}(q^\dagger\mathcal{L}^{(0)}p) \neq 0$ for spherical tensor operators $p,q \in \{T_{l,m}\}$; this basis is chosen for the representation since both $\Gamma$ and $\mathcal{H}$ matrices are block diagonal in the projection quantum number $m$.  The first three columns represent each line of \autoref{eq:HIS} and confirm the behavior described above, where in \autoref{fig:Lmat}b spin coupling terms are absent and in \autoref{fig:Lmat}c the field $\mathbf{B}_0$ is averaged to zero.

\autoref{fig:Lmat} highlights the different spherical symmetry of the spin-spin and spin-field couplings.  The $J_{IS}$ coupling is invariant to global rotation of quantization axis and therefore overall angular momentum cannot be changed, so in the first column only basis operators with $l_p=l_q$ and $m_p=m_q$ are connected.  The Zeeman interaction, however, may connect operators with $m_p=m_q$ or $m_p=m_q\pm1$.   Matrices for dipole-dipole relaxation and cross-correlated\cite{Ghose1999MolPhys96} random field relaxation in the motional narrowing limit represented in the two right-hand columns are also diagonal in $l$ and $m$, except for dipole-dipole relaxation under XY4($I$) where rotation symmetry is broken slightly.  Due to the different symmetries, contributions to $\mathcal{L}^{(0)}$ do not share a common eigenbasis.  This means that eigenvalues of $\Gamma$ generally do not equate to overall relaxation rates, and moreover that overall relaxation of the system depends on the pulse sequence used.  In this example, XY4($I$) averages out cross-correlated relaxation.

\begin{figure}%
	\includegraphics[width=8.99cm]{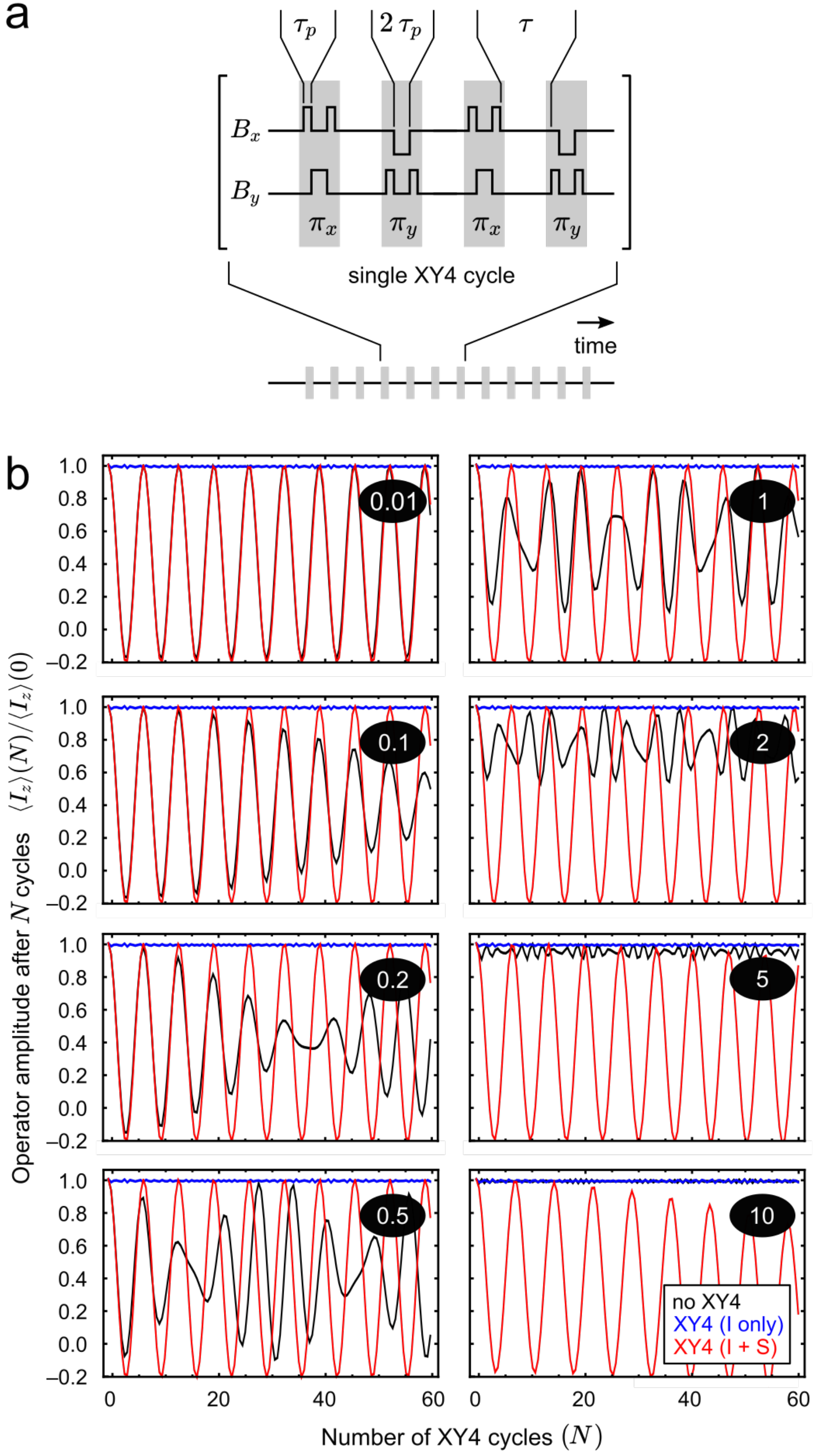}
	\caption{Numerical simulation of ultralow-field nuclear spin symmetry switching: (a) Pulse sequence comprising a periodic train of dc rotation pulses of duration $4\tau_p$ and effective flip angle $\pi$, which are separated by a time delay $\tau$.  Alternate pulses are applied along the $x$ and $y$ axes.   (b) simulated trajectories of $I$-spin polarization along the $z$ axis during the sequence shown in (a).  Representative values $J=2.5$ \si{\hertz}, and $\tau=0.01$ \si{\second} are used.  Each plot shows trajectories for a different dimensionless magnetic field $\textbf{B}'_0\equiv\{0,0,|(\gamma_I-\gamma_S)B_z/(2\pi J_{IS})|\}$ in the crossover regime between strong and weak internuclear coupling, where the z component is marked in the black ovals.  Three curves are shown, which correspond to cases of (black, $\tau_p=0$ s) no applied pulses, where a clear change in dynamics is observed through the crossover, (blue, $\tau_p=46$ \si{\micro\second}) XY4 pulses on the $I$ spin only, resulting in decoupling and a stationary $I$-spin polarization, plus (red, $\tau_p=7\times46$ \si{\micro\second}) XY4 pulses on both spins to impose zero-field conditions, resulting in polarization oscillations.}% 
	\label{fig:XY4deuterium}
\end{figure}

Experimentally speaking, in a near-zero magnetic field one cannot perform spin-selective rotations such as those in XY4($I$) using conventional rf pulses, since differences in Larmor frequency are small compared to rf pulse bandwidths\cite{Kreis1988JCP89}.  Simple sequences of dc pulses are used instead.  For $I$ = \textsuperscript{1}H and $S$ = \textsuperscript{2}H the nominal operation $R_x^{(I)}(\pi)R_x^{(S)}(0)$ is approximated using a composite of three dc pulses (\autoref{fig:XY4deuterium}a) that rotates the $I$ spin first by $\pi/2$ about the $x$ axis, then $\pi$ about $y$ and finally $\pi/2$ about $x$, since for the ratio $u=\gamma_S/\gamma_I \approx 0.154$ in this case it is found that $R_x^{(I)}(\pi/2)R_y^{(I)}(\pi)R_x^{(I)}(\pi/2) = R_x^{(I)}(\pi)$ and $R_x^{(S)}(u\pi/2)R_y^{(S)}(u\pi)R_x^{(S)}(u\pi/2) \approx R_x^{(S)}(0)$. 

If pulse flip angles are multiplied by $v$ to approximate $\pi/2$ and $\pi$ rotations on the $S$ spin, the same dc pulse sequence can be used to produce $R_x^{(I)}(\pi)R_x^{(S)}(\pi)$ transformations, provided that $v$ is odd and $(1-|vu|)$ is less than the error-compensation bandwidth of the overall rotation\cite{LevittEMRcompositepulses}.  Here $v=7$ yields $R_x^{(I)}(v\pi/2)R_y^{(I)}(v\pi)R_x^{(I)}(v\pi/2) = R_x^{(I)}(\pi)$ and $R_x^{(S)}(vu\pi/2)R_y^{(S)}(vu\pi)R_x^{(S)}(vu\pi/2) \approx R_x^{(S)}(\pi)$.  Consequently, XY4($I$) and XY4($I$+$S$) can be implemented using dc pulses.  

Numerical simulations demonstrate that both versions of XY4 are highly accurate when applied using dc pulses of finite length.  For example, \autoref{fig:XY4deuterium}b shows the evolution of a \textsuperscript{1}H-\textsuperscript{2}H spin pair starting from $\rho(0) = I_z$ versus number of XY4 cycles applied.  Each panel shows a different condition of a dimensionless magnetic field $\mathbf{B}'_0\equiv\{0,0,|(\gamma_I-\gamma_S)B_z/(2\pi J_{IS})|\}$ between strong and weak coupling regimes.  The strong coupling regime is marked by $|\mathbf{B}'_0| < 1$ and corresponds to the plots in the left-hand column, while weak coupling is marked by $|\mathbf{B}'_0| \gg 1$.  In the simulation, $J_{IS} = 2.5$ Hz is chosen as a typical coupling between \textsuperscript{1}H and \textsuperscript{2}H nuclei attached to adjacent carbon atoms in an aliphatic chain, thus the range of $B_z$ is \SI{0.7}{\nano\tesla} to \SI{700}{\nano\tesla}, and the interval between pulse centers, $\tau = $ \SI{0.01}{\second} is much shorter than $1/J_{IS} = $ \SI{0.4}{\second}.   Differences between the curves are immediately seen.  In the case of free evolution nearest to zero field (black curves, $|\mathbf{B_0}'|=0.1$), the expectation value of $I_z$ oscillates at a frequency $J_{IS}$ corresponding to polarization exchange back and forth with $S$.  At the high-field end, the expectation value is roughly constant because the secular part of the Hamiltonian commutes with $I_z$. % (\gtext{the non-secular part is responsible for the small-amplitude oscillation}).  

In agreement with the analysis presented earlier, all XY4($I$) and XY4($I+S$) trajectories are invariant across the range of $B_z$ and converge to the free-evolution system behavior in the high- and zero-field limits.  The simulations therefore confirm that the error $[1-\gamma(^1\rm H)/(7\gamma(^2\rm H))] \approx 0.06$ lie within the inversion bandwidth of the pulses and that the phase cycle compensates for residual imperfections. Experimental benchmarks for these sequences are presented in the Supporting Information.

\begin{figure}%
	\includegraphics[width=9cm]{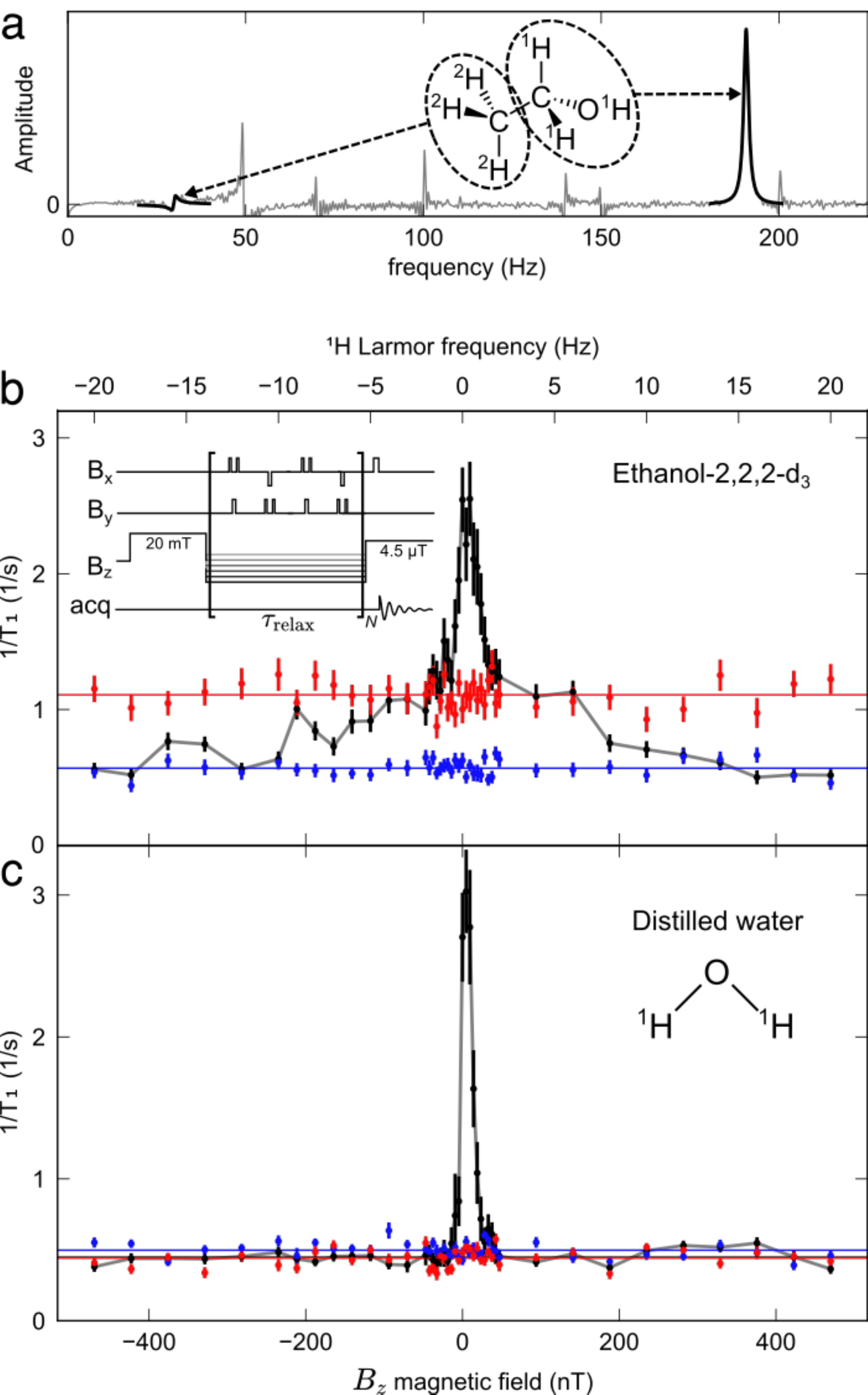}
	\caption{Manipulation of \textsuperscript{1}H nuclear spin evolution near zero field in pure CD\textsubscript{3}CH\textsubscript{2}OH and pure H\textsubscript{2}O: (a) Optical-magnetometer-detected NMR spectra of the CH\textsubscript{2}OH and CD\textsubscript{3} subsystems at \SI{4.46}{\micro\tesla} showing single unresolved resonances for \textsuperscript{1}H and \textsuperscript{2}H. Respective Larmor frequencies are 191 Hz and 29 Hz and full-widths-at-half-maximum are \SI{1.63(4)}{Hz} and \SI{1.05(23)}{Hz}.  Other peaks including 50 Hz and harmonics are due to background technical noise in the laboratory; (b) Fitted monoexponential relaxation rates for the \textsuperscript{1}H signal in CD\textsubscript{3}CH\textsubscript{2}OH versus $B_z$ under conditions of (black) free evolution, (blue) XY4(\textsuperscript{1}H) and (red) XY4(\textsuperscript{1}H+\textsuperscript{2}H) using $v=7$; (c) Fitted monoexponential relaxation rates for the \textsuperscript{1}H signal in H\textsubscript{2}O.  The same color coding denotes the pulse sequence.
	}% 
	\label{fig:d3ethanol}
\end{figure}

\noindent Experimental -- anhydrous trideuterated ethanol (CD\textsubscript{3}CH\textsubscript{2}OH, 99.9~$\%$ pure liquid) is chosen as a material to experimentally investigate scalar relaxation near zero field, albeit a more challenging system to analyze than the $IS$ spin pair since it contains six spins in total.  The two \textsuperscript{1}H nuclei in the CH\textsubscript{2} group are scalar coupled to \textsuperscript{2}H spins in the CD\textsubscript{3} with a coupling constant of $^3J_{\rm HD}=1.2$ Hz\cite{Tayler2019JMR298} and are also coupled with the third \textsuperscript{1}H nucleus (OH) with $^2J_{\rm HH}=5.4$ Hz\cite{DeVience2021CPC22}.  In the high-field limit therefore, all \textsuperscript{1}H spins are regarded as coupling strongly to one another and weakly to \textsuperscript{2}H.  The resulting NMR spectrum contains one broad unresolved line per spin species (\autoref{fig:d3ethanol}a) \cite{Tayler2019JMR298}.

Since \textsuperscript{1}H spins couple to one another strongly, the relaxation of their total magnetization $\propto T_{10}^{(I)}$ is quantified by a single exponential time constant,  $T_1$(\textsuperscript{1}H).  The dependence of $T_1$(\textsuperscript{1}H) on $B_z$ is measured using a fast-field cycling protocol and presented in \autoref{fig:d3ethanol}b.  The method involves first pre-magnetizing the spins in a $B_z=20$ mT magnetic field, allowing them to relax for a time $\tau_{\rm relax}$ near $B_z=0$, and finally returning to $B_z= $\SI{4.46}{\micro\tesla} where a free-precession signal is detected, of amplitude proportional to $\exp(-\tau_{\rm relax}/T_1({\rm\textsuperscript{1}H}))$.

During $\tau_{\rm relax}$ spins may evolve freely or under the constraint of XY4 pulse trains ($\tau_p=$ \SI{46}{\micro\second} for $I$, $\tau_p=$ \SI{324}{\micro\second} for $I$+$S$ using $v=7$, $\tau$ = \SI{0.01}{\second}).  Qualitatively the dispersion of relaxation rates in \autoref{fig:d3ethanol}b is consistent with coherent averaging theory where for both XY4($I$) (blue) and XY4($I$+$S$) (red) curves the rate is constant across the range \SI{-0.5}{\micro\tesla} $<B_z<$ \SI{+0.5}{\micro\tesla}, where $\mathcal{L}^{(0)}$ should not depend on $\mathbf{B}_0$.  The rate difference between XY4($I$) and XY4($I$+$S$) is then attributed to scalar relaxation and amounts to \SI{0.5(1)}{s^{-1}} in the ethanol system -- in other words, that the \textsuperscript{1}H relaxation time is almost doubled when the coupling to \textsuperscript{2}H is the dominant coherent interaction.  For comparison, \autoref{fig:d3ethanol}c shows relaxation rates for \textsuperscript{1}H in a sample of pure water (\textsuperscript{1}H\textsubscript{2}O).  As H\textsubscript{2}O contains only \textsuperscript{1}H spins, XY4($I$) and XY4($I$+$S$) dynamical decoupling sequences are effectively the same and no rate differences are observed. 

There also occurs below $|B_z| = $ \SI{50}{\nano\tesla} a large increase in the apparent relaxation rate for both CD\textsubscript{3}CH\textsubscript{2}OH and H\textsubscript{2}O in the case of free evolution (black curves in \autoref{fig:d3ethanol}b and \autoref{fig:d3ethanol}c).  The behavior is attributed to non-secular components $B_x$ and $B_y$ of the background field of order \SI{10}{\nano\tesla}, which switch the axis of Larmor precession away from $B_z$ during $\tau_{\rm relax}$.  This hypothesis is confirmed by manual adjustment of the background field to set $B_x=B_y=0$ and eliminate the feature (see Supporting Information).  However, such a process is extremely time consuming and due to field drifts must be performed regularly.  The absence of the feature for XY4 data indicate that dynamical decoupling is a faster and more reliable method to obviate effects of residual fields, and is consistent with the absence of $\mathbf{B}_0$ terms from the lowest-order average Liouvillian.

\begin{figure}
	\includegraphics[width=10cm]{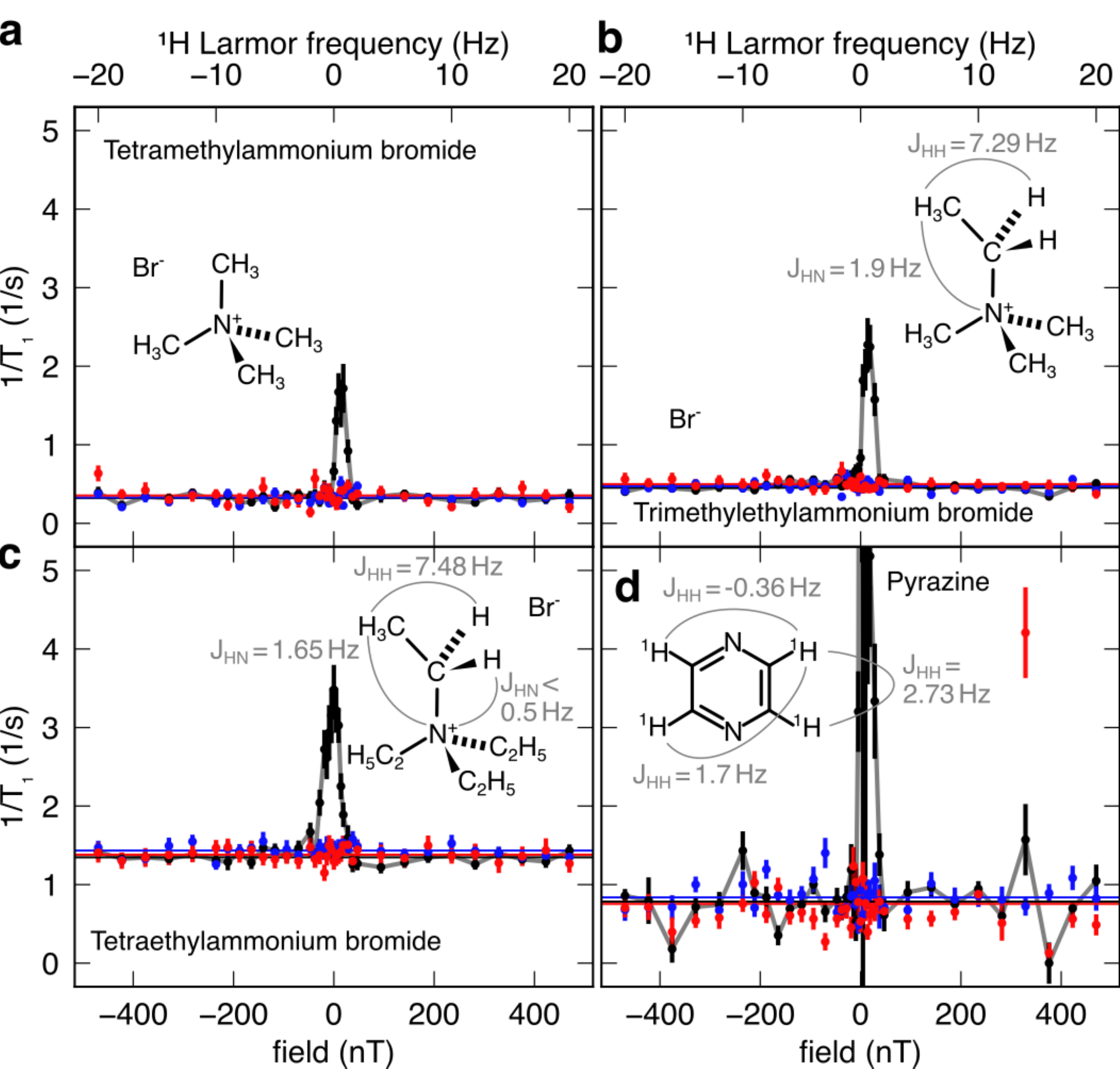}
	\caption{Fitted monoexponential relaxation rates for the \textsuperscript{1}H signal of \textsuperscript{14}N compounds solvated in D\textsubscript{2}O: (a) tetramethylammonium bromide, N(CH\textsubscript{3})\textsubscript{4}Br; (b) ethyltrimethylammonium bromide, N(CH\textsubscript{3})\textsubscript{3}(CH\textsubscript{2}CH\textsubscript{3})Br; (c) tetraethylammonium bromide, N(CH\textsubscript{2}CH\textsubscript{3})\textsubscript{4}Br; (d) pyrazine, C\textsubscript{4}H\textsubscript{4}N\textsubscript{2}.  Colors represent the dynamical decoupling conditions: (black) no pulses, free evolution; (blue) XY4(\textsuperscript{1}H); (red) XY4(\textsuperscript{1}H+\textsuperscript{14}N) using $v=13$.
	}
	\label{fig:netma}
\end{figure}

Finally, nitrogen-14 is another spin-1 species abundant in organic compounds such as amines, peptides and ammonium salts.  In many of these compounds it is found that protic exchange at the nitrogen atom provides efficient pathways for spin decoherence\cite{Barskiy2019NComms10,Birchall2020ChemComm56,Shchepin2018JPCC122}.  However, in non-exchanging systems the decoherence mechanisms due to \textsuperscript{14}N-\textsuperscript{1}H couplings appear to be much weaker, even negligible.  This is shown by the \textsuperscript{1}H relaxation dispersions plotted in \autoref{fig:netma}a-d where for a selection of \textsuperscript{14}N-containing compounds in D\textsubscript{2}O solution ($\approx$ 2 M), there is no measurable difference between the free evolution, XY4(\textsuperscript{1}H) or XY4(\textsuperscript{1}H+\textsuperscript{14}N) scenarios despite heteronuclear J couplings of a few Hz\cite{Bullock1963JCP38,Lehn1965JCP43,Porter1984CHC3}, similar in magnitude to those found in [2,2,2-d\textsubscript{3}]-ethanol.  The hypothesis of weak heteronuclear coupling is supported by a lack of dependence on the point symmetry of the nitrogen atom, which influences the magnitude of the \textsuperscript{14}N nuclear quadrupole moment and thus \textsuperscript{14}N relaxation rate.  It could be argued that the lower number of spin states to which \textsuperscript{1}H nuclei may couple, $(2S+1)=3$ in mono-\textsuperscript{14}N compounds (\autoref{fig:netma}a-c) compared with $(2S+1)^3=27$ in [2,2,2-d\textsubscript{3}]-ethanol, limits heteronuclear decoherence effects to below 100 nT total field.  Yet, even in this central region the XY4 sequences should distinguish the \textsuperscript{14}N-induced decoherence from effects of residual $B_x$ and $B_y$ components, and neither curve in \autoref{fig:netma} shows significant field dependence.

In summary, accurate and spin-species-selective $\pi$-pulse trains made possible using XY4 phase cycling and dc composite pulses can be used to modify the effective Liouvillian in heteronuclear spin systems at ultralow magnetic field.  These enable symmetry switching between effective zero-field and high-field coupling regimes, an application of which is to suppress spin relaxation that occurs via couplings to quadrupolar species, such as deuterium. A timely application of the concepts introduced is to the efficient production of parahydrogen-enhanced molecular imaging probes\cite{Hovener2018AN57} where the singlet spin order of parahydrogen is converted via chemical interaction into in-phase magnetization on the probe molecule.  Transfer to long-lived \textsuperscript{13}C magnetization can be achieved using parahydrogen close to zero fields, but during polarization transfer it is desired to minimize the number of connected spins.  Only a few molecules satisfy the latter condition naturally, e.g.\ fumarate\cite{Knecht2021PNAS118}. Proton-deuterium decoupling, therefore, may offer other molecules to be polarized via similar ultralow-field routes, without deleterious effects\cite{Svyatova2021SciRep,Salnikov2018JPCC122}.

\section*{Experimental Methods}

\subsection*{Sample Preparation}
All liquid samples studied in this work were contained in disposable glass vials (12 mm o.d., 20 mm length, 1.8 mL internal volume, 8-425 thread) sealed with a silicone septum and finger-tight polypropylene screw cap.  Samples were obtained commercially (Sigma Aldrich) except for NEtMe\textsubscript{3}Br, which was synthesized as described in the Supporting Information. 

\subsection*{Ultralow-Field NMR Relaxometry}
NMR signals from the samples were measured at $B_z=$ \SI{4.46}{\micro\tesla} inside a magnetic shield following initial prepolarization of the nuclear spins in a parallel field of $\sim$\SI{20}{\milli\tesla} and then a dc $\pi/2$ pulse applied along $x$.  An atomic magnetometer of sensitivity \SI{8}{\femto\tesla\per\sqrt\hertz} adjacent to the NMR sample was used to detect the time-dependent $y$ component of nuclear magnetization after the dc pulse.  The magnetometer signal was Fourier transformed and a Lorentzian line shape was fitted to the peak at the \textsuperscript{1}H Larmor frequency, \SI{190}{\hertz}.

In between prepolarization and the $\pi/2$ pulse, fast field cycling was used to expose the spins to ultralow fields $|B_z|<$ \SI{500}{\nano\tesla}.  A detailed description of the procedure is given here\cite{Bodenstedt2021NatComm}.  Additionally, to produce spatially homogeneous magnetic fields along $x$ and $y$ axes for the XY4 sequences, two saddle coils of 13 cm diameter, 13 cm length and relative orientation \SI{90}{\degree} were added inside the central layer of the magnetic shield (6 turns/loop, 29 American wire gauge, \SI{80}{\micro\tesla\per\ampere}).  Current to the coils was supplied from a dc power supply (Conrad PS2403D) and switched using a 2-channel dc motor driver chip (Toshiba TB6612FNG, \SI{1}{\micro\second} pulse rise and fall time).  The amplitude of pulsed field used was $|\mathbf{B}_1|=$ \SI{128}{\micro\tesla}, thus a $\pi/2$ rotation on \textsuperscript{1}H was produced using $\tau_{p}=$\SI{44}{\micro\second}.  Finally, a microcontroller (ARM Cortex-M4 MK64FX512, 120 MHz CPU, \SI{2}{\micro\second} time base) was used to control the logic lines of the motor driver, the field cycling and the acquisition of the magnetometer signal.  By varying the time spent at ultralow field the fitted amplitudes of the NMR signals produced a decay curve, which at each field was fitted to a monoexponential decay function with time constant $T_1$.

%%%%%%%%%%%%%%%%%%%%%%%%%%%%%%%%%%%%%%%%%%%%%%%%%%%%%%%%%%%%%%%%%%%%%
%% The "Acknowledgement" section can be given in all manuscript
%% classes.  This should be given within the "acknowledgement"
%% environment, which will make the correct section or running title.
%%%%%%%%%%%%%%%%%%%%%%%%%%%%%%%%%%%%%%%%%%%%%%%%%%%%%%%%%%%%%%%%%%%%%
\begin{acknowledgement}
The work described is funded by: 
EU H2020 Marie Sk{\l}odowska-Curie Actions project ITN ZULF-NMR  (Grant Agreement No. 766402); 
Spanish MINECO project OCARINA (PGC2018-097056-B-I00 project funded by MCIN/ AEI /10.13039/501100011033/ FEDER ``A way to make Europe''); 
the Severo Ochoa program (Grant No. SEV-2015-0522); 
Generalitat de Catalunya through the CERCA program; 
Ag\`{e}ncia de Gesti\'{o} d'Ajuts Universitaris i de Recerca Grant No. 2017-SGR-1354;  Secretaria d'Universitats i Recerca del Departament d'Empresa i Coneixement de la Generalitat de Catalunya, co-funded by the European Union Regional Development Fund within the ERDF Operational Program of Catalunya (project QuantumCat, ref. 001-P-001644);
Fundaci\'{o} Privada Cellex; 
Fundaci\'{o} Mir-Puig; 
MCD Tayler acknowledges financial support through the Junior Leader Postdoctoral Fellowship Programme from ``La Caixa'' Banking Foundation (project LCF/BQ/PI19/11690021). 
\end{acknowledgement}

%%%%%%%%%%%%%%%%%%%%%%%%%%%%%%%%%%%%%%%%%%%%%%%%%%%%%%%%%%%%%%%%%%%%%
%% The same is true for Supporting Information, which should use the
%% suppinfo environment.
%%%%%%%%%%%%%%%%%%%%%%%%%%%%%%%%%%%%%%%%%%%%%%%%%%%%%%%%%%%%%%%%%%%%%

\section*{Author Contributions}
MCD Tayler proposed the study.  
S Bodenstedt and D Moll prepared the samples.  
S Bodenstedt built the experimental apparatus, measured and analyzed the experimental data and together with MCD Tayler made the theoretical interpretation.  
MCD Tayler wrote the manuscript with input from all authors.  
All authors reviewed the manuscript and suggested improvements.  
MCD Tayler and MW Mitchell supervised the overall research effort. 

\section*{Competing Interests}
There are no competing interests to declare.

\begin{suppinfo}
Experimental benchmarking of dc composite-pulse sequences,
Shimming procedure,
Synthesis of N(C\textsubscript{2}H\textsubscript{5})(CH\textsubscript{3})\textsubscript{3}Br,
Average Liouvillian theory calculations and numerical simulation of spin dynamics trajectories (Mathematica version 11.1). % Link to electronic material is automatically added by publisher.
\end{suppinfo}

%%%%%%%%%%%%%%%%%%%%%%%%%%%%%%%%%%%%%%%%%%%%%%%%%%%%%%%%%%%%%%%%%%%%%
%% The appropriate \bibliography command should be placed here.
%% Notice that the class file automatically sets \bibliographystyle
%% and also names the section correctly.
%%%%%%%%%%%%%%%%%%%%%%%%%%%%%%%%%%%%%%%%%%%%%%%%%%%%%%%%%%%%%%%%%%%%%


\begin{thebibliography}{99} \label{sec:TeXbooks}

\bibitem{Souza2012PhilTransRoyalSocA} Souza, A.\,M.; \'Alvarez, G.\,A.; Suter, D. Robust dynamical decoupling. \textit{Phil. Trans. A} \textbf{2012}, \textit{370} (1976), 4748--4769. DOI: 10.1098/rsta.2011.0355
%https://doi.org/10.1098/rsta.2011.0355

\bibitem{Levitt1992PRL69} Levitt, M.\,H.; Di Bari, L. Steady state in magnetic resonance pulse experiments. \textit{Phys. Rev. Lett.} \textbf{1992}, \textit{69}(21), 3124--3127. DOI:10.1103/physrevlett.69.3124

\bibitem{Levitt1994BMR16} Levitt, M.\,H.; Di Bari, L. The homogeneous master equation and the manipulation of relaxation networks. \textit{Bull. Magn. Reson.} \textbf{1994}, \textit{16} (1--2), 94--114.
% https://citeseerx.ist.psu.edu/viewdoc/download?doi=10.1.1.173.1586&rep=rep1&type=pdf

\bibitem{Ghose1999MolPhys96} Ghose R.; Eykyn, T.; Bodenhausen, G. Average Liouvillian theory revisited: cross correlated relaxation between chemical shift anisotropy and dipolar couplings in the rotating frame in nuclear magnetic resonance. \textit{Mol. Phys.} \textbf{1999}, \textit{96} (8), 1281--1288. DOI: 10.1080/00268979909483072

\bibitem{Carravetta2004JACS126}
Carravetta, M.; Levitt, M.\,H. Long-lived nuclear spin states in high-field solution NMR, \textit{J. Am. Chem. Soc.} \textbf{2004}, \textit{126} (20), 6228--6229. DOI: 10.1021/ja0490931
% https://doi.org/10.1021/ja0490931

\bibitem{Kimmichbook} \textit{Field-cycling NMR relaxometry: instrumentation, model theories and applications}; Kimmich, R., Ed.; The Royal Society of Chemistry, Oxford, 2018. %ISBN 978-1-78801-154-9.

\bibitem{Kimmich1979BMR1} Kimmich, R. Field cycling in NMR relaxation spectroscopy: applications in biological, chemical and polymer physics, \textit{Bull. Magn. Reson.} \textbf{1979}, \textit{1} (4), 195--218.

\bibitem{Kimmich2004PNMRS} Kimmich, R. and Anoardo, E., Field-cycling NMR relaxometry, \textit{Prog. Nucl. Magn. Reson. Spectrosc.} \textbf{2004}, \textit{44} (47) , 257--320. DOI: 10.1002/chin.200447278

\bibitem{Levitt2020Pileiobook} Levitt, M.H. Long-lived states in nuclear magnetic resonance: an overview.  In \textit{Long-lived spin order: theory and applications}; Royal Society of Chemistry, 2020; pp 1--32. %.  ISBN 978-1-78801-568-4.
%https://doi.org/10.1039/9781788019972-00001 

\bibitem{VINOGRADOV2007JMR176}
Vinogradov, E.; Grant, A.\,K. Long-lived states in solution NMR: selection rules for intramolecular dipolar relaxation in low magnetic fields. \textit{J. Magn. Reson.} \textbf{2007}, \textit{188} (1), 176--182. DOI: 10.1016/j.jmr.2007.05.015
%https://doi.org/10.1016/j.jmr.2007.05.015}

\bibitem{Karabanov2009JCP131} Karabanov, A.\,A.; Bretschneider, C.; K\"{o}ckenberger, W. Symmetries of the master equation and long-lived states of nuclear spins. \textit{J. Chem. Phys.} \textbf{2009}, \textit{131} (204105), 204105, 1--10. DOI:  10.1063/1.3265852
% https://doi.org/10.1063/1.3265852

\bibitem{Emondts2014PRL112} Emondts, M.; Ledbetter, M.\,P.; Pustelny, S.; Theis, T.; Patton, B.; Blanchard, J.\,W.; Butler, M.\,C.; Budker, D.; Pines, A. Long-lived heteronuclear spin-singlet states in liquids at a zero magnetic field. \textit{Phys. Rev. Lett.} \textbf{2014}, \textit{112} (077601), 1--5. DOI: 10.1103/PhysRevLett.112.077601

\bibitem{Tayler2018JMR297} Tayler, M.\,C.\,D.; Ward-Williams, J.; Gladden, L.\,F. NMR relaxation in porous materials at zero and ultralow magnetic fields. \textit{J. Magn. Reson.} \textbf{2018}, \textit{297}, 1--8. DOI: 10.1016/j.jmr.2018.09.014% https://doi.org/10.1016/j.jmr.2018.09.014.

\bibitem{BudkerOpticalMagnetometrybook} \textit{Optical magnetometry}; Budker, D.,  Jackson Kimball, D.\,F., Eds.; Cambridge University Press, 2013. %, ISBN: 1107010357.

\bibitem{Tayler2017RSI} Tayler, M.\,C.\,D. ; Theis, T.; Sjolander, T.\,F.;  Blanchard, J.\,W.; Kentner, A.; Pustelny, S; Pines, A.; Budker, D. Instrumentation for nuclear magnetic resonance in zero and ultralow magnetic field. \textit{Rev. Sci. Instrum.} \textbf{2017}, \textit{88} (9). DOI: 10.1063/1.5003347

\bibitem{Zhukov2018PCCP} Zhukov, I.\,V. et al. Field-cycling NMR experiments in an ultra-wide magnetic field range: relaxation and coherent polarization transfer. \textit{Phys. Chem. Chem. Phys.} \textbf{2018}, \textit{20}, 12396--12405. DOI: 10.1039/c7cp08529 % https://10.1039/c7cp08529j.

\bibitem{Kiryutin2020Pileiobook} Kiryutin, A.\,S.; Zhukov, I.\,V.; Yurkovskaya, A.\,V.; Budker, D.; Ivanov, K.\,L. Singlet order in heteronuclear spin systems.  In \textit{Long-lived spin order: theory and applications}, Pileio, G., Ed.; Royal Society of Chemistry, 2020. % ISBN 978-1-78801-568-4.
%https://doi.org/10.1039/9781788019972-00418 

\bibitem{Blanchard2016emagres} Blanchard, J.\,W.; Budker, D. Zero- to ultralow-field NMR. \textit{eMagRes} \textbf{2016}, \textit{5} (3), 1395--1409. DOI: 10.1002/9780470034590.emrstm1369

\bibitem{Blanchard2013JACS} Blanchard, J.\,W.; Ledbetter, M.\,P.; Theis, T.; Butler, M.; Budker, D.; Pines, A. High-resolution zero-field NMR J-spectroscopy of aromatic compounds. \textit{J. Am. Chem. Soc.} \textbf{2013}. \textit{135} (9), 3607--3612. DOI: 10.1021/ja312239v

\bibitem{Goldman2006CRC9} Goldman, M.; J{\'o}hannesson, H.; Axelsson, O.; Karlsson, M. Design and implementation of \textsuperscript{13}C hyper polarization from para-hydrogen, for new MRI contrast agents. \textit{Compt. Rend. Chim.} \textbf{2006}, \textit{9} (3--4), 357--363. DOI: 10.1016/j.crci.2005.05.010
% https://doi.org/10.1016/j.crci.2005.05.010

\bibitem{Cavallari2015JPCB119} Cavallari, E.; Carrera, C.; Boi, T.; Aime, S.; Reineri, F. Effects of magnetic field cycle on the polarization transfer from parahydrogen to heteronuclei through long-range J-Couplings. \textit{J. Phys. Chem B} \textbf{2015}, \textit{119} (31), 10035--10041. DOI: 10.1021/acs.jpcb.5b06222
% https://doi.org/10.1021/acs.jpcb.5b06222

\bibitem{Stewart2018JMR296} Stewart, N.\,J.; Kumeta, H.; Tomohiro, M.; Hashimoto, T.; Hatae, N.; Matsumoto, S. Long-range heteronuclear J-coupling constants in esters: implications for 13C metabolic MRI by side-arm parahydrogen-induced polarization. \textit{J. Magn. Reson.} \textbf{2018},  \textit{296}, 85--92. DOI: 10.1016/j.jmr.2018.08.009
%https://doi.org/10.1016/j.jmr.2018.08.009}

\bibitem{Hartwig2011JCP135} Hartwig, S.; Voigt, J.; Scheer, H.-J.; Albrecht, H.-H.; Burghoff, M.; Trahms, L. Nuclear magnetic relaxation in water revisited. \textit{J. Chem. Phys.} \textbf{2011}, \textit{135} (054201), 1--5. DOI: 10.1063/1.3623024

\bibitem{Barskiy2017CPC} Barskiy, D.\,A.;  Shchepin, R.\,V.;  Tanner, C.\,P.\,N., Colell, J.\,F.\,P.; Goodson, B.\,M.; Theis, T.;  Warren, W.\,S.; Chekmenev, E.\,Y. The absence of quadrupolar nuclei facilitates efficient \textsuperscript{13}C hyperpolarization via reversible exchange with parahydrogen. \textit{ChemPhysChem} \textbf{2017}, \textit{18} (12), 1493--1498. DOI: 10.1002/cphc.201700416

\bibitem{Tayler2019JMR298} Tayler, M.\,C.\,D.; Gladden, L.\,F. Scalar relaxation of NMR transitions at ultralow magnetic field. \textit{J. Magn. Reson.} \textbf{2019}, \textit{298}, 101--106. DOI: 10.1016/j.jmr.2018.11.012 % https://doi.org/10.1016/j.jmr.2018.11.012.

\bibitem{Sjolander2017JCPletters8} Sjolander, T.\,F.; Tayler, M.\,C.\,D.; Kentner, A.; Budker, D.; Pines, A. \textsuperscript{13}C-decoupled J-coupling spectroscopy using two-dimensional nuclear magnetic resonance at zero field. \textit{J. Phys. Chem. Lett.} \textbf{2017}, \textit{8} (7), 1512--1516. DOI: 10.1021/acs.jpclett.7b00349
% https://doi.org/10.1021/acs.jpclett.7b00349

\bibitem{DeVience2021CPC22} DeVience, S.\,J.; Greer, M.; Mandal, S; Rosen, M.\,S. Homonuclear J-coupling spectroscopy at low magnetic fields using spin-lock induced crossing. \textit{ChemPhysChem} \textbf{2021}, \textbf{22}(22), 2128--2137. DOI: 10.1002/cphc.202100162
%https://doi.org/10.1002/cphc.202100162

\bibitem{Kreis1988JCP89} Kreis, R.; Thomas, A.; Studer, W.; Ernst, R.\,R. Low frequency pulse excitation in zero field magnetic resonance. \textit{J. Chem. Phys.} \textbf{1988}. \textbf{89} (11), 6623-6635. DOI: 10.1063/1.455384  %https://doi.org/10.1063/1.455384 

\bibitem{LevittEMRcompositepulses} Levitt, M.\,H. Composite pulses. In \textit{eMagRes}; Harris, R.\,K.,Wasylishen, R.\,L., Eds.; 2007. DOI: 10.1002/9780470034590.emrstm0086 %https://doi.org/10.1002/9780470034590.emrstm0086

\bibitem{Lee1987JMR75} Lee, C.\,J.; Suter, D.; Pines, A. Theory of multiple-pulse NMR at low and zero fields, \textit{J. Magn. Reson.} \textbf{1987}. \textit{75}, 110-124.

\bibitem{Bodenstedt2021NatComm} Bodenstedt, S.; Mitchell, M.\,W.; Tayler; M.\,C.\,D., Fast-field-cycling, ultralow field nuclear magnetic relaxation dispersion, \textit{Nat. Commun.} \textbf{2021}, \textit{12} (4041). DOI: 10.1038/s41467-021-24248-9 % https://doi.org/10.1038/s41467-021-2424

\bibitem{Ganssle2014ANIE53} Ganssle, P.\,J.; et al. Ultra-low-field NMR relaxation and diffusion measurements using an optical magnetometer, \textit{Angew. Chem. Intl. Edn.} \textbf{2014}, \textit{53} (37), 9766--9770. DOI: 10.1002/anie.201403416

\bibitem{Savukov2005PRL} Savukov, I.\,M.; Romalis, M.\,V., NMR detection with an atomic magnetometer, \textit{Phys. Rev. Lett.} \textbf{2005}, \textit{94} (123001), 1--4. DOI: 10.1103/PhysRevLett.94.123001 

\bibitem{Budker2007NatPhys3} Budker, D.; Romalis, M.\,V. Optical magnetometry, \textit{Nat. Phys.} \textbf{2007}, \textbf{3}, 227--234. DOI: 10.1038/nphys566

\bibitem{Barskiy2019NComms10} Barskiy, D.\,A.; Tayler, M.\,C.\,D.; Marco-Rius, I.; et al. Zero-field nuclear magnetic resonance of chemically exchanging systems, \textit{Nat. Commun.} \textbf{2019}, \textit{10} (3002). DOI: 10.1038/s41467-019-10787-9 %https://doi.org/10.1038/s41467-019-10787-9

\bibitem{Birchall2020ChemComm56} Birchall, J.\,R.; Kabir, M.\,S.\,H.; Salnikov, O.\,G.; Chukanov, N.\,V.; Svyatova, A.; Kovtunov, K.\,V.; Koptyug, I.\,V.; Gelovani, J.\,G.; Goodson, B.\,M.; Phamg, W.; Chekmenev, E.\,Y. Quantifying the effects of quadrupolar sinks via \textsuperscript{15}N relaxation dynamics in metronidazoles hyperpolarized via SABRE-SHEATH. \textit{Chem. Comm.} \textbf{2020}, \textit{56} (64), 9098--9101. DOI: 10.1039/D0CC03994B
%https://doi.org/10.1039/D0CC03994B

\bibitem{Shchepin2018JPCC122} Shchepin, R.\,V.; Jaigirdar, L.; Chekmenev, E.\,Y. Spin-lattice relaxation of hyperpolarized metronidazole in signal amplification by reversible exchange in micro-tesla fields. \textit{J. Phys. Chem. C} \textbf{2018}, \textit{122} (9), 4984--4996. DOI: 10.1021/acs.jpcc.8b00283
%https://doi.org/10.1021/acs.jpcc.8b00283

\bibitem{Bullock1963JCP38}  Bullock, E.; Tuck, D.\,G.; Woodhouse, E.\,J.
Unusual spin--spin couplings in NMR spectra of alkyl ammonium salts. \textit{J. Chem. Phys.} \textbf{1963}, \textit{38} (9), 2318--2318. DOI: 10.1063/1.1733981 
%https://doi.org/10.1063/1.1733981 

\bibitem{Lehn1965JCP43} Lehn, J.-M.; Franck--Neunamm, M.; Nuclear spin-spin interactions. V. \textsuperscript{1}H-\textsuperscript{14}N spin-spin coupling and quadrupolar relaxation in quaternary ammonium salts. \textit{J. Chem. Phys.} \textbf{1965}, \textit{43} (4), 1421--1422. DOI: 10.1063/1.1696936
%https://doi.org/10.1063/1.1696936

\bibitem{Porter1984CHC3} Porter, A.\,E.\,A. Pyrazines and their benzo derivatives. \textit{Compr. Heterocyc. Chem.} \textbf{1984}, \textit{3}, 157--197.

\bibitem{Hovener2018AN57} H\"ovener, J.-B.; Pravdivtsev, A.\,N.; Kidd, B.; Bowers, C.\,R.; Gl\"{o}ggler, S.; Kovtunov, K.\,V.; Plaumann, M.; Katz-Brull, R.; Buckenmaier, K.; Jerschow, A.; Reineri, F.; Theis, T.; Shchepin, R.\,V.; Wagner, S.; Bhattacharya, P.; Zacharias, N.\,M.; Chekmenev, E.\,Y.; Parahydrogen-based Hyperpolarization for Biomedicine, \textit{Angew. Chem. Int. Ed. Engl.} \textbf{2018}, \textit{57}, 11140--11162.

\bibitem{Knecht2021PNAS118} Knecht, S.; Blanchard, J.\,W.; Barskiy, D.; Cavallari, E.; Dagys, L.; Van Dyke, E; Tsukanov, M.; Bliemel, B.; M{\"u}nnemann, K.; Aime, S.; Reineri, F.; Levitt, M.\,H.; Buntkowsky, G.; Pines, A.; Bl{\"u}mler, P.; Budker, D.; Eills, J.; Rapid hyperpolarization and purification of the metabolite fumarate in aqueous solution, \textit{Proc. Natl. Acad. Sci. USA}, \textbf{2021}, \textit{118 (13)}, e2025383118.
%https://doi.org/10.1073/pnas.2025383118

\bibitem{Svyatova2021SciRep} Svyatova, A.; Kozinenko, V.\,P.; Chukanov, N.\,V.; Burueva, D.\,B.; Chekmenev, E.\,Y.; Chen, Y.-W.; Hwang, D.\,W.; Kovtunov, K.\,V.; Koptyug, I.\,V.; PHIP hyperpolarized [1-\textsuperscript{13}C]pyruvate and [1-\textsuperscript{13}C]acetate esters via PH-INEPT polarization transfer monitored by \textsuperscript{13}C NMR and MRI, \textit{Sci. Rep.} \textbf{2021}, \textit{11}, 5646. %https://doi.org/10.1038/s41598-021-85136-2

\bibitem{Salnikov2018JPCC122} Salnikov, O.\,G.; Shchepin, R.\,V.; Chukanov, N.\,V.; Jaigirdar, L.; Pham, W.; Kovtunov, K.\,V.; Koptyug, I.\,V.; Chekmenev, E.\,Y.; Effects of Deuteration of C-Enriched Phospholactate on Efficiency of Parahydrogen-Induced Polarization by Magnetic Field Cycling. \textit{J. Phys. Chem. C -- Nanomaterials and Interfaces} \textbf{2018}, \textit{122}, 24740-24749. %PMID 31447960 DOI: 10.1021/Acs.Jpcc.8B07365

\end{thebibliography}
\end{document}